\documentclass[epj]{svjour}
\usepackage{graphics}
\begin{document}
\title{Influence  of density dependent symmetry energy on Elliptical flow.}

\author{ Karan Singh Vinayak \and Suneel Kumar} 

\offprints{suneel.kumar@thapar.edu}                        

\institute{School of Physics and Materials Science, Thapar University, Patiala-147004, Punjab (India)}
\date{Received: date / Revised version: date}
%

\abstract{The effect of density dependent symmetry energy is studied on elliptical flow is studied using isospin-dependent quantum molecular dynamics model(IQMD). We have used the reduced isospin-dependent cross-section with hard(H) equation of state to study the sensitivity of elliptical flow towards symmetry energy in the energy range of 50 - 1000 MeV/nucleon. The elliptical flow becomes zero at a particular energy termed as transition energy. A systematic effort has been made to pin down the transition energy for the density dependent symmetry energy.}

%

\maketitle
\section{Introduction}
\label{intro}

 Information about the nuclear matter under the extreme conditions of temperature and density and the role of symmetry energy under these conditions is still a topic of crucial importance in the present day nuclear physics research. Quite good progress has been made with the help of heavy-ion collisions to study the nature of equation of state \cite{ad690,rk47}. The form and strength of symmetry energy, and the study of different observables at sub-nuclear densities with respect to the symmetry energy is still a challenging task in the intermediate energy heavy ion collisions(HIC's). Among different observables, collective flow and its various forms enjoys a special status, due to its sensitivity towards the model ingredients that define a equation of state.\\

 Recently, many efforts have been made to study the collective flow both theoretically and experimentally in heavy-ion collisions \cite{31,41,51,510,61,71,81,var}. The collective flow is a motion characterized by the space-momentum correlations of dynamic origin. The directed flow and elliptical flow are  the two parameters available in the heavy-ion collisions to study the collective flow. The directed flow seems to be constrained only along the reaction plane which is due to the bounce-off of the compressed matter. However, the elliptical flow is more suited to study the collective flow which is a squeeze out of the spectator matter out of the reaction plane \cite{81,11,var}. The highly stopped and compressed nuclear matter around the mid-rapidity region is seen directly in the squeeze out. Generally, the elliptical flow is the difference between the major and minor axes. It describes the eccentricity of an ellipse like distribution. The elliptical flow is defined by the second-order Fourier coefficient from the azimuthal distribution of detected particles at mid-rapidity.\\

Mathematically,

\begin{equation}
\frac{dN}{d\phi}~=~p_o(1+2v_{1}cos\phi+2v_{2}cos2\phi).
\end{equation}

Here, $\phi$ is the azimuthal angle between the transverse momentum of the particle and reaction plane. The parameters $\langle Cos2\phi \rangle$ of elliptical flow depends upon the complex interplay between the expansion, rotation and shadowing of the spectators, apart from the incident energy. The particular energy at  which elliptical flow vanishes is termed as the transition energy. The elliptical flow arises due to the orthogonal asymmetry in the configuration space and rescattering.\\

 One needs to understand that the high density symmetry energy term can be probed from the isospin effects in HIC's. The elliptical flow, which provides us an opportunity to study the pressure that is generated very early during the reaction \cite{11}, is highly sensitive towards the symmetry energy\cite{81,z,25,chan054}. The studies based on the Transport model calculations concluded, that the squeezed out of nucleons with high transverse momenta perpendicular to the reaction plane, i.e. elliptical flow, has probably the highest sensitivity towards the symmetry energy\cite{chan054}.\\

 Recently, an attempt was made by S. Kumar el. al.\cite{81} to correlate the elliptical flow with isospin content of a reaction. Various studies \cite{81,25} concluded the elliptical flow as a powerful probe for the symmetry energy. Unfortunately, all the calculations were silent about the density dependence of the symmetry energy. The symmetry energy which is the difference of the energy per nucleon between pure neutron matter and symmetric neutron matter is taken as 32 MeV corresponding to the normal nuclear matter density i.e. $\rho$ = 0.16 $fm^{-3}$\cite{chan054,sakshi}. This understanding does not remain valid as one goes away from the normal nuclear matter density and symmetric nuclear matter\cite{chan054,sakshi}.  The knowledge of the density dependence of the symmetry energy and its equation of state for isospin asymmetric nuclear matter is of crucial importance, to study the structure of the systems as diverse as the neutron rich nuclei and the neutron stars. A large number of studies have been performed on the density dependence of the symmetry energy in the recent past \cite{chan054,dvsh0246,zhan}. The equation below gives us the most extensively used theoretical parametrization of how the symmetry energy varies against $\rho$ \cite{chan054,dvsh0246,zhan}.\\

\begin{eqnarray}
E(\rho)=E(\rho_o)(\rho/\rho_o)^{\gamma},
\end{eqnarray}

The larger(smaller) value of the constant $\gamma$ corresponds to stiff(soft) density dependence of the symmetry energy\cite{chan054,dvsh0246,zhan}.
 Experimentally, symmetry energy is not a directly measurable quantity and has to be extracted from the observables which can shed light on symmetry energy. Some authors from the FOPI data concluded that the symmetry energy at supra-densities is very soft \cite{fopi1}. However, others concluded an opposite findings\cite{fopi2}. Recently, Shetty et.al.\cite{dvsh0246}, and Tsang et. al.\cite{tsang}, concluded the density dependence of the symmetry energy with $\gamma$  = 0.4 - 1.05.\\

In a recent communication, it was advocated that at high baryon densities, large symmetry energy repulsion affects the transverse in-plane flow \cite{sakshi} of a neutron rich matter in heavy-ion collisions. A large sensitivity for the transverse in-plane flow was detected by assuming the symmetry potential as a function of density.
However, it could be of the interest to carry out a meaningful investigation for elliptical flow, by including the density dependence of the symmetry energy. No such study is reported in the literature so far. Our present aim is to pin down the density dependent symmetry energy via transition energy.\\ 

The present study is carried out within the framework of the isospin-dependent quantum molecular dynamics(IQMD) model \cite{5}. Section II describes the model in brief. Section III discusses the results, and Section IV summarizes the results.


\section{ISOSPIN-dependent QUANTUM MOLECULAR DYNAMICS (IQMD) MODEL}
\label{sec:2}

The IQMD model \cite{5}, which is an improved version of the QMD model \cite{3,6} developed by J. Aichelin and coworkers and then improved by Puri and coworkers has been applied to explain various phenomenon such as collective flow, disappearance of flow, fragmentation \& elliptical flow successfully. 
The isospin degree of freedom enters into the calculations via symmetry potential, cross-sections and
Coulomb interactions\cite{5}.
The details about the elastic and inelastic cross-sections
for proton-proton and neutron-neutron collisions can be found in Ref.\cite{5}.\\

In IQMD model, the nucleons of target and projectile
interact via two and three-body Skyrme forces, Yukawa potential and Coulomb interactions. 
In addition to the use of explicit charge states of all baryons and mesons, a symmetry potential between 
protons and neutrons corresponding to the Bethe- Weizsacker mass formula has been included.\\
The hadrons propagate using classical Hamilton equations of motion:
\begin{eqnarray}
\frac{d\vec{r_i}}{dt}~=~\frac{d\it{\langle~H~\rangle}}{d{p_i}}~~;~~\frac{d\vec{p_i}}{dt}~=~-\frac{d\it{\langle~H~\rangle}}{d{r_i}}
\end{eqnarray}
with
\begin{eqnarray}
\langle~H~\rangle&=&\langle~T~\rangle+\langle~V~\rangle\nonumber\\
&=&\sum_{i}\frac{p_i^2}{2m_i}+
\sum_i \sum_{j > i}\int f_{i}(\vec{r},\vec{p},t)V^{\it ij}({\vec{r}^\prime,\vec{r}})\nonumber\\
& &\times f_j(\vec{r}^\prime,\vec{p}^\prime,t)d\vec{r}d\vec{r}^\prime d\vec{p}d\vec{p}^\prime .
\end{eqnarray}
 The baryon-baryon potential $V^{ij}$, in the above relation, reads as:
\begin{eqnarray}
V^{ij}(\vec{r}^\prime -\vec{r})&=&V^{ij}_{Skyrme}+V^{ij}_{Yukawa}+V^{ij}_{Coul}+V^{ij}_{sym}+V_{mdi}\nonumber\\
&=&\left(t_{1}\delta(\vec{r}^\prime -\vec{r})+t_{2}\delta(\vec{r}^\prime -\vec{r})\rho^{\gamma-1}
\left(\frac{\vec{r}^\prime +\vec{r}}{2}\right)\right)\nonumber\\
& & +~t_{3}\frac{exp(|\vec{r}^\prime-\vec{r}|/\mu)}{(|\vec{r}^\prime-\vec{r}|/\mu)}
~+~\frac{Z_{i}Z_{j}e^{2}}{|\vec{r}^\prime -\vec{r}|}\nonumber\\
& &+t_{6}\frac{1}{\varrho_0}T_3^{i}T_3^{j}\delta(\vec{r_i}^\prime -\vec{r_j})\nonumber\\
& &+t_{7}ln^{2}[t_{8}(\vec{p_i}^\prime-\vec{p})^{2}+1]
\delta(\vec{r_{i}^\prime}-\vec{r}).
\label{s1}
\end{eqnarray}

Here $Z_i$ and $Z_j$ denote the charges of $i^{th}$ and $j^{th}$ baryon, and $T_3^i$, $T_3^j$ are their respective $T_3$
components (i.e. 1/2 for protons and -1/2 for neutrons).
The parameters $\mu$ and $t_1,.....,t_6$ are adjusted to the real part of the nucleonic optical potential. 
For the density
dependence of nucleon optical potential, standard Skyrme-type parameterization is employed.
The potential part resulting from the convolution of the distribution function 
with the Skyrme interactions $V_{\it Skyrme}$ reads as :
\begin{equation}
{\it V}_{Skyrme}~=~\alpha\left(\frac{\rho_{int}}{\rho_{0}}\right)+\beta\left(\frac{\rho_{int}}{\rho_{0}}\right)^{\gamma}
~~\cdot
\end{equation}
The two of the three parameters of equation of state are determined by demanding that at normal nuclear matter density,
 the binding energy should be equal to 16 MeV. 
\begin{equation}
\kappa~=~9\rho^{2}\frac{\partial^{2}}{\partial\rho^{2}}\left(\frac{E}{A}\right)~~\cdot
\end{equation}
$\kappa$ is the compressibility factor. The different values of compressibility give rise to soft(S) and hard(H) equations of state(EOS). It is worth mentioning that as shown by Puri and coworkers, Skyrme 
forces are very successful in the analysis of low energy phenomena such as fusion, fission and cluster-radioactivity, where nuclear potential plays an important role.\cite{8}. The momentum dependent interactions can be incorporated by parametrizing the momentum dependence of real part of optical potential\cite{mom}.

Many studies concluded the matter to be soft, whereas many more believe the matter to be harder in
nature \cite{harder1,harder2}. As noted\cite{hard}, elliptical flow is unaffected by
the choice of equation of state.
For the present analysis, a hard (H) equation of state
has been employed along with reduced isospin-dependent cross-section(0.9 of $\sigma_{NN}$).\\

\section{Results and Discussion}
\label{sec:3}

We here perform a complete systematic study for the reaction of $Au^{179}_{97}+Au^{179}_{97}$, using different parametrizations of the density dependence of symmetry energy. In addition to that, the reactions of $Sn^{124}_{50}+Sn^{124}_{50}$ and $Xe^{124}_{54}+Xe^{124}_{54}$ are also simulated to acknowledge the effect of isospin content on the elliptical flow. The phase apace generated by the IQMD model has been subjected to clusterization using minimum spanning tree method[MST] \cite{8}  and analysis packages\cite{6}. The MST method binds two nucleons in a fragment if their distance is less than 4 fm. However, in recent times many effective and proficient,  algorithms are also available in the literature. The quantum molecular dynamics(QMD) generated phase space analysed through one  such algorithm simulated annealing clusterization algorithm (SACA)\cite{saca} tends to solve the long standing problem of QMD method. The reactions are followed until the elliptical flow saturates i.e. at 200 fm/c.\\

\begin{figure}
\resizebox{0.55\textwidth}{!}{%
  \includegraphics{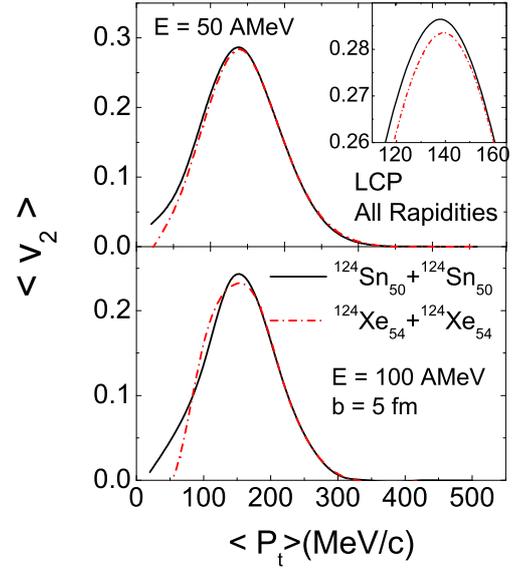}
}
\caption{Transverse momentum dependence of elliptical flow, summed over entire rapidity distribution, for LCP's at 50(top) and 100 MeV/nucleon(bottom), respectively. Both reactions have same mass number but different N/Z ratio.}
\label{fig:1}       
\end{figure}

To study the effect of symmetry energy. We display in fig. 1, the reactions of $Xe^{124}_{54}+Xe^{124}_{54}$  and $Sn^{124}_{50}+Sn^{124}_{50}$ under the same conditions for LCP's.
 Interestingly, the elliptical flow seems to be more sensitive for the N/Z ratio at incident energy of 100 MeV/nucleon. The trends observed through our calculations shows that irrespective of the same mass, the weaker squeeze-out flow is observed in case of the neutron rich system. These findings are also in accordance with the findings in ref.\cite{81,zhang}.    
At the incident energy of 50 MeV/nucleon, the isospin does not play a larger role. In such case, binary collisions are rare and mean field dominates the dynamics of the reaction. On the other hand, around 100 MeV/nucleon, the mean field and binary collisions both contribute towards the dynamics of the reaction. In such a case larger effect of isospin can be observed. Therefore, it could be further interesting to carry out the study of elliptical flow for the various forms of the density dependence of the symmetry energy.\\

\begin{figure}
\resizebox{0.58\textwidth}{!}{%
  \includegraphics{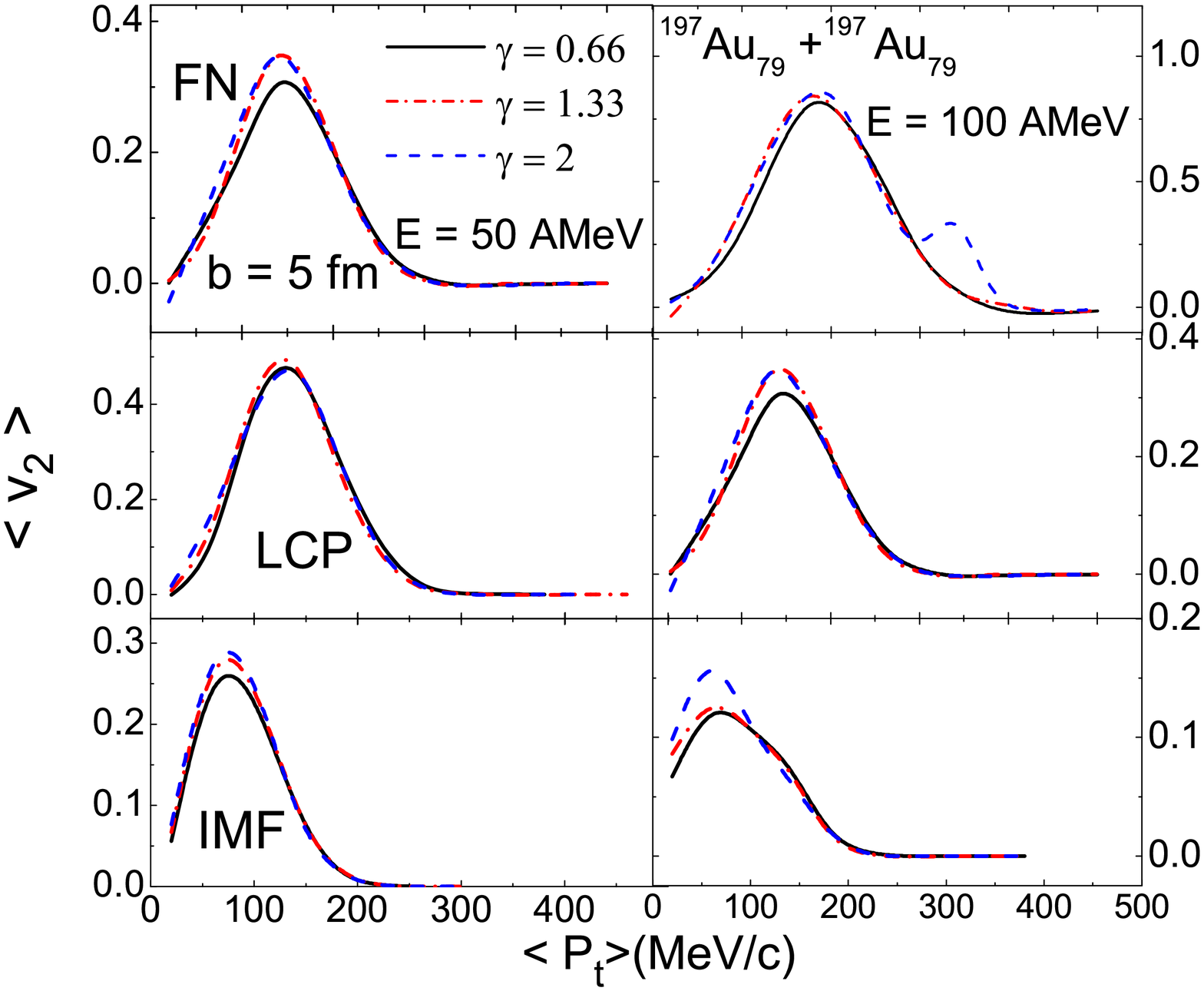}
}
\caption{Transverse momentum dependence of the elliptical flow,
summed over the entire rapidity distribution, at b = 5 fm for the system $Au^{179}_{97}+Au^{179}_{97}$  for different forms of the density dependence of the symmetry energy, at 50 (left) and 100 (right) MeV/nucleon. The top, middle, and bottom panels represent the free nucleons (FN's),
light charged particles (LCP's), and intermediate mass fragments(IMF's) respectively.}
\label{fig:2}       
\end{figure}

As the symmetry energy acts directly on the the squeeze out nucleons, that are mostly from the high density region, formed during an early stage of the reaction and are not affected by the spectator nucleons\cite{chan054}. For neutron rich system, $Au^{179}_{97}+Au^{179}_{97}$ large role of symmetry energy can be expected with reference to the elliptical flow. The study of elliptical flow by constraining the density dependence of the symmetry energy can give us better understanding of nucleon-nucleon interaction and its extrapolation to the behaviour of symmetry energy in neutron-rich nuclei.  In fig. 2, we display the, final state elliptical flow, at different incident energies, for the free particles (upper panel), light charged particles (LCP's) [$2 \leq A \leq 4 $](middle) and intermediate mass fragments [IMF's] ($5\leq A \leq A_{tot}/6$](lower panel) as a function of transverse momentum $(P_{t})$ for the various forms of the density dependence of the symmetry energy i.e. $\gamma$ = 0.66, 1.33 and 2, respectively. The pressure created during a collision can be revealed through the elliptical flow, with respect to transverse momenta. \\
\begin{figure}
\resizebox{0.60\textwidth}{!}{%
  \includegraphics{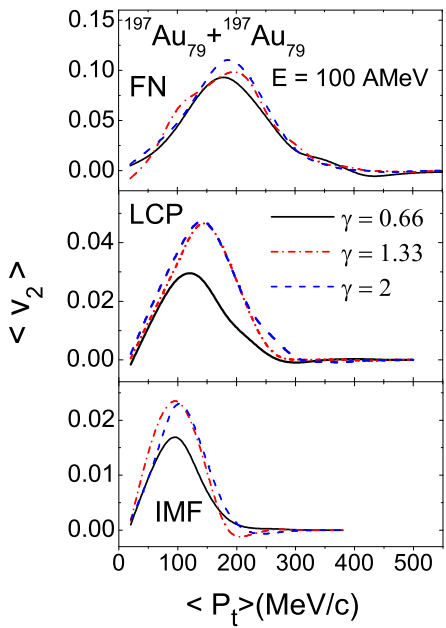}
}
\caption{Transverse momentum dependence of the elliptical flow,
for mid rapidity $(\mid y \mid = \frac{y_{c.m.}}{y_{c.m.}} \leq 0.1)$, at b = 5 fm for the system $Au^{179}_{97}+Au^{179}_{97}$  for different forms of the density dependence of the symmetry energy, at 100 (right) MeV/nucleon. The top, middle, and bottom panels represent the free nucleons (FN's),
light charged particles (LCP's), and intermediate mass fragments(IMF's) respectively.}
\label{fig:3}       
\end{figure}

This elliptical flow is integrated over the entire rapidity range. Gaussian-type, behaviour is observed for all the forms of the density dependence of the symmetry energy i.e. for the different values of gamma.\\

The effect of symmetry energy is clearly visible in the figure. The elliptical flow seems to be sensitive towards the various forms of the density dependence of the symmetry energy. The trends observed through our simulations shows the weaker squeeze out flow for the larger values of gamma. The very stiff form of symmetry energy i.e.  $\gamma = 2$, corresponds to the maximum positive value of elliptical flow. Indeed, the very stiff form of symmetry energy i.e. $\gamma = 1.33$ and 2 does not seems to affect the squeeze out flow on a large scale.\\

The isospin effect basically generates from the mid-rapidity region or in other words from the participant zone. To study the effect of symmetry energy, we parametrized the $\langle v_2 \rangle$ as a function of $ \langle P_t \rangle$ at mid-rapidity$(|y| = \frac{y_{c.m.}}{y_{c.m.}} \leq 0.1)$ in fig. 3. We display the FN's(top panel), LCP's(middle panel) and IMF's(bottom panel) for the various forms of the density dependence of symmetry. We see that the squeeze out is affected more in case of LCP's as compared to other fragments. Although, the heavier fragments have weaker senstivity towards the symmetry energy.\\

The different forms of the symmetry energy tends to affect the elliptical flow considerably. The most interesting aspect in this case can be the variation of the transition energy $(E_{Trans})$ with different forms of the density dependent symmetry energy. The incident energy at which elliptical flow vanishes is termed as the transition energy. Although, the variation in the squeeze out of nuclear matter due to the various forms of the symmetry energy can be understood with the transition energy.\\

\begin{figure}
\resizebox{0.52\textwidth}{!}{%
  \includegraphics{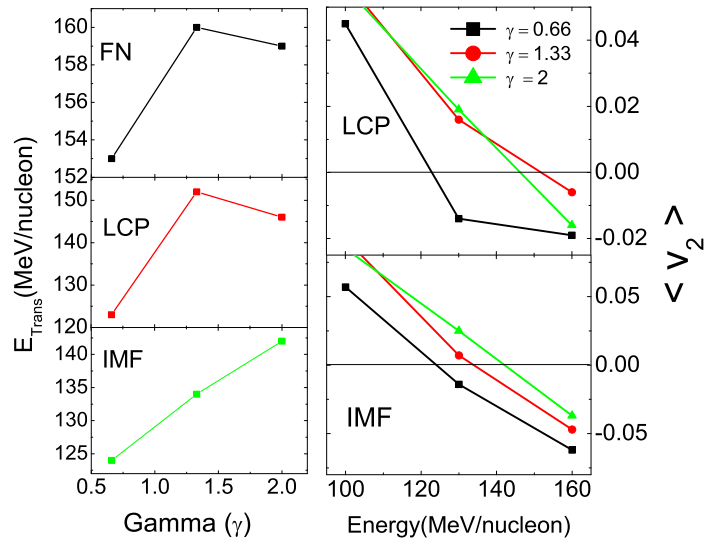}
}
\caption{Same as in fig.3, but for the transition energy dependence of gamma($\gamma$).}
\label{fig:4}       
\end{figure}

In fig.4, we parametrized the transition energy with the stiffness factor $\gamma$.
The transition energy tends to increase with the increase in value of gamma. The light charged particles however tends to behave in a different manner as compared to the intermediate mass fragments. As, the FN's and LCP's originate from the participant zone. The larger sensitivity of free nucleons and light charged particles tends to increase the transition energy upto a certain level. The  $(E_{Trans})$ tends to decrease for very stiff form of the symmetry energy. Although, the IMF's are being generated solely from the spectator zone. The transition energy for the IMF's increases with $\gamma$. As the IMF's being heavy particles move slowly as compared to FN's and LCP's. The symmetry energy tends to play a role at later stages in case of heavy fragments(IMF's). Therefore, in case of FN's and LCP's  transition energy tends to decrease at very stiff density dependence of symmetry energy, which is not in case of IMF's.   The elliptical flow which is generated  due to the compression in the participant zone, tends to give different behaviour for the heavy fragments(generated from spectator zone) as compared to the light mass fragments(generated from the participant zone).\\

\begin{figure}
\resizebox{0.50\textwidth}{!}{%
  \includegraphics{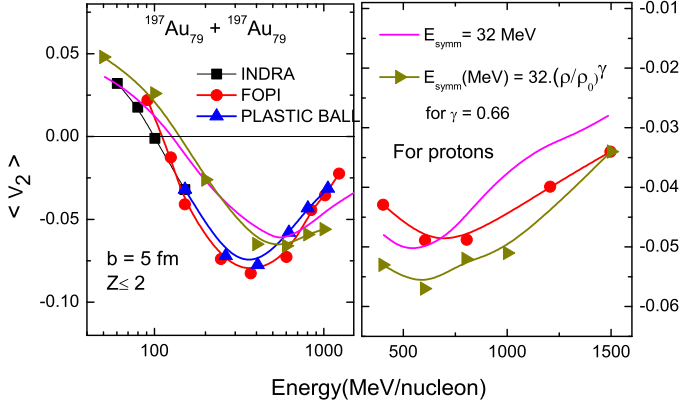}
}
\caption{Variation of the elliptical flow, summed
over the entire transverse momentum, with beam energy at $ |y| =
\frac{y_{c.m.}}{y_{beam}} \leq 0.1$ for the  reaction $Au^{179}_{97}+Au^{179}_{97}$ . Here a comparison
is shown with the constant form of symmetry energy and the density dependent symmetry energy for gamma = 0.66, with the experimental findings of the INDRA, FOPI, and
PLASTIC BALL Collaborations\cite{41,51,510}}
\label{fig:5}       
\end{figure}

In fig. 5, we show elliptical flow, $\langle v_2 \rangle$ at mid rapidity $( |y| =
\frac{y_{c.m.}}{y_{beam}} \leq 0.1)$ for Z $\leq$ 2 (left panel) and for  protons(right panel) as a function of the incident energy. The rapidity cut is in accordance with the experimental findings. We compare the theoretical results with constant form of the symmetry energy and the density dependent symmetry energy with $\gamma = 0.66$, with the experimental data extracted by INDRA, FOPI, and
PLASTIC BALL Collaborations\cite{41,51,510}. The elliptical flow shows a transition from in-plane to out-of-plane with increase in incident energy.
The elliptical flow is found to become more negative with the increase in incident energy. This is because the mean field, which contributes to the formation of a rotating compound system, becomes less important, and the collective expansion process based on nucleon-nucleon scattering starts predominant. The out of plane emission decreases again toward the higher incident energies. This happens due to the faster movement of the spectator matter after $ \langle v_2 \rangle$ reaches a maximal negative value. The density dependent symmetry energy tends to give the more squeeze out of the nuclear matter as compared to the constant form.\\

 For $\gamma = 0.66$, the symmetry energy tends to play larger role across the high density region. The repulsion produced due to the symmetry energy tends to give the larger squeeze out flow, which is evolved due to the pressure generated in the participant zone. The trends obtained through the density dependent symmetry energy  for $\gamma = 0.66$ are in more agreement with the experimental data as compared to the constant form of the symmetry energy. However, no such effect is observed in case of the low incident energies. At low incident energies the density of the system is not so large to see a huge variation in the symmetry energy. As in the case of higher incident energies of the order of 800-1000 MeV/nucleon, the density of the system is 2-3 times the normal nuclear matter density. As the mid rapidity region corresponds to the participant zone. The high density achieved in the participant zone, due to target and projectile collisions, tends to produce the effect of density dependent symmetry energy. The larger squeeze-out is also observed in case of protons.\\

%
\section{Conclusion}
\label{sec:4}

We have checked the sensitivity of elliptical flow for the various forms of the  symmetry energies. We found that the elliptical flow is highly sensitive towards the different forms of the density dependent symmetry energy.  Transition energy tends to give different behaviour for the lighter particles as compared to the heavy fragments. However, larger squeeze out is observed with the stiff density dependence of symmetry energy$(\gamma =0.66)$ as compared to the full symmetry energy strength i.e. 32 MeV.


%

\end{document}